\documentclass[aps,pre,twocolumn,superscriptaddress,showpacs]{revtex4}
\usepackage{amssymb}
\usepackage{epsfig}
\usepackage{amsmath}
\usepackage{times}

\begin{document}

\title{Oscillations of complex networks}
\author{Xingang Wang}
\affiliation{Department of Physics, National University of Singapore, Singapore 117542}
\affiliation{Beijing-Hong Kong-Singapore Joint Centre for Nonlinear \& Complex Systems
(Singapore), National University of Singapore, Kent Ridge, Singapore 119260}
\author{Ying-Cheng Lai}
\affiliation{Department of Electrical Engineering, Department of Physics and Astronomy,
Arizona State University, Tempe, Arizona 85287, USA}
\author{Choy Heng Lai}
\affiliation{Department of Physics, National University of Singapore, Singapore 117542}
\affiliation{Beijing-Hong Kong-Singapore Joint Centre for Nonlinear \& Complex Systems
(Singapore), National University of Singapore, Kent Ridge, Singapore 119260}

\begin{abstract}
A complex network processing information or physical flows is usually
characterized by a number of macroscopic quantities such as the diameter and
the betweenness centrality. An issue of significant theoretical and
practical interest is how such a network responds to sudden changes caused
by attacks or disturbances. By introducing a model to address this issue, we
find that, for a finite-capacity network, perturbations can cause the
network to \emph{oscillate} persistently in the sense that the
characterizing quantities vary periodically or randomly with time. We
provide a theoretical estimate of the critical capacity-parameter value for
the onset of the network oscillation. The finding is expected to have broad
implications as it suggests that complex networks may be structurally highly
dynamic.
\end{abstract}

\date{\today }
\maketitle

The response of a complex network to sudden changes such as intentional
attacks, random failures, or abnormal load increase, has been of great
interest \cite%
{AJB:2000,CEAH:2000,CNSW:2000,Watts:2002,HK:2002,MGP:2002,Holme:2002,ML:2002,GGKK:2003,Motter:2004}
since the discoveries of the small-world \cite{WS:1998} and the scale-free
\cite{BA:1999} topologies. The issue is particularly relevant for scale-free
networks that are characterized by a power-law degree distribution. For such
a network, generically there exists a small set of nodes with degrees
significantly higher than those of the rest of the nodes. A scale-free
network is thus robust against random failures, but it is vulnerable to
intentional attacks \cite{AJB:2000}. This is particularly so when dynamics
on the network is taken into account, which can lead to catastrophic
breakdown of the network via the cascading process \cite{HK:2002,ML:2002}
even when the attack is on a single node. A basic assumption underlying the
phenomenon of cascading breakdown is that a node fails if the load exceeds
its capacity. As a result, the load of the failed node has to be transferred
to other nodes, which causes more nodes to fail, and so on, leading to a
cascade of failures that can eventually disintegrate the network.

There are situations in complex networks where overload does not necessarily
lead to failures. For instance, in the Internet, when the number of
information-carrying packets arriving at a node exceeds what it can handle,
traffic congestion occurs. That is, overload of a node can lead to the
waiting of packets but not to the failure of the node. As a result of the
congestion, traffic detour becomes necessary in the sense that any optimal
routes for new packets on the network try to avoid the congested nodes. This
is equivalent to a change in the ``weights'' (to be defined more precisely
below) of the congested nodes and, consequently, to changes in the
macroscopic characterizing quantities of the network. This situation usually
does not occur when the network is in a normal operational state, but it
becomes likely when sudden disturbances, such as an attack or an abrupt
large load increase, occur. A question is then whether the network can
recover after a finite amount of time, in the sense that its characterizing
quantities restore to their original values.

In this Letter, we study a class of weighted scale-free networks,
incorporating a feasible traffic-flow protocol, to address the above
question. In the absence of any perturbations, the network is assumed to
operate in its \textquotedblleft normal\textquotedblright\ state so that its
macroscopic characterizing quantities are constants. We find that, after a
large perturbation, the network can indeed recover but only for large node
capacities. When the node capacities are not significantly higher than their
loads in the normal state, a surprising phenomenon arises: the macroscopic
quantities of the network are never able to return to their unperturbed
values but instead, they exhibit persistent oscillations. In this sense we
say the \emph{network oscillates}. More remarkably, as the node capacities
are decreased, both periodic and random oscillations can occur. The striking
feature is that the oscillations, periodic or random, are caused solely by
the interplay between the network topology and the traffic-flow protocol,
regardless of the network parameters such as the degree distribution and the
overall load fluctuations. For fixed network parameters, the oscillations
exist despite of the explicit form of the local node dynamics, given the
simple rule that it \textquotedblleft holds\textquotedblright\ and causes
the traffic to wait when overloaded. This may have wide implications to many
traffic problems. For instance, it can provide an alternative explanation,
from the dynamical point of view, for the recently observed random
oscillations in real Internet traffic flow \cite{VB:2000,GRHA:2005} and give
some insights to the self-similar oscillations of the traffic flux observed
in WWW \cite{LTWW:1994,CB:1997}.

We begin by constructing a scale-free network of $N$ nodes using the
standard growth and preferential-attachment mechanism \cite{BA:1999}. We
next define the node capacity by using the model in Ref. \cite{ML:2002},
\begin{equation}  \label{eq:capacity_definition}
C_i = (1 + \alpha) L_i(0),
\end{equation}
where $L_i(0)$ is the initial load on node $i$, which is approximately the
load in a normal operational state (free of traffic congestion), and $\alpha
> 0$ is the \emph{capacity} parameter. The load $L_i$ can be conveniently
chosen to be the betweenness \cite{Newman:2001,GKK:2001}, which is the total
number of optimal paths \cite{Optimal_Path} between all pairs of nodes
passing through node $i$. To define an optimal path at time $t$, say at this
time the weights associated with node $i$ and with node $j$ are $w_i(t)$ and
$w_j(t)$ (to be defined below according to the degree of traffic
congestion), respectively, where there is a direct link $l_{ij}$ between the
two nodes. The weight of the link is then $d_{ij}(t) = [w_i(t) + w_j(t)]/2$.
Given a pair of nodes, one packet generating and another receiving, the
optimal path is the one which minimizes the sum of all weights $d_{ij}$ of
links that constitute the path. Finally, we define a traffic protocol on the
network by assuming that, at each time step, one packet is to be
communicated between any pair of nodes. There are thus $N(N-1)/2$ packets to
be transported across the whole network at any time. When a packet is
generated, its destination and the optimal path that the packet is going to
travel toward it are determined.

In a computer or a communication network, a meaningful quantity to
characterize a link is the time required to transfer a data packet through
this link. When the traffic flow is free, it takes one time unit for a node
to transport a packet. When congestion occurs, it may take a substantially
long time for a packet to pass through a node and hence any link from this
node. For instance, suppose at time $t$ there are $J_i(t)$ packets at node $%
i $, where $J_i(t) > C_i$. Since the node can process $C_i$ packets at any
time, the waiting time for a packet at the end of the queue is $1 + %
\mbox{int}[J_i(t)/C_i]$, where $\mbox{int}[\cdot]$ is the integer part of
the fraction in the square bracket. These considerations lead to the
following definition of \emph{instantaneous} weight for node $i$:
\begin{equation}  \label{eq:weight_definition}
w_i(t) = 1 + \mbox{int}[\frac{J_i(t)}{C_i}], \ \ \mbox{for} \ \ i = 1,
\ldots, N,
\end{equation}
from which the instantaneous weights for any link in the network and hence a
set of instantaneous optimal paths can be calculated accordingly. For free
traffic flow on the network, we have $J_i(t) < C_i$ and hence $w_i(t) = 1$
so that the network is non-weighted. In this case, the optimal path reduces
to the shortest path.

The above model of traffic dynamics on a weighted network allows us to
investigate the response of the network to perturbations in a systematic
way. In particular, since the node capacities are the key to the occurrence
of traffic congestion, it is meaningful to choose the capacity parameter $%
\alpha $ in Eq. (\ref{eq:capacity_definition}) as a bifurcation parameter.
To apply perturbation, we locate the node with the largest betweenness $%
B_{max}$ in the network and generate a large number of packets,
say ten times of $B_{max}$, at time $t=0$. The network is then
allowed to relax according to our model. Initially, because of the
congestion at the largest-betweenness node caused by the
perturbation, its weight assumes a large value. As a result, there
is a high probability that the optimal paths originally passing
through this node change routes. This can lead to a sudden
increase in the network diameter, which is the average of all
optimal paths. As time goes by, the congestion will cascade to
other nodes which adopt the detoured optimal paths and, as a
result, the diameter will increase very quickly and reaches its
maximum at some instant. During this process the congestion
situation is released at the attacking node while it get worse at
nodes that paths detour to. After this, the network begins to
"absorb" the congestions according to the load tolerance, i.e. the
capacity parameter, and the recovery process starts. During this
process the network congestion is gradually released and the
diameter is expected to be decreasing. The same processes apply to
other macroscopic characterizing quantities of the network, such
as the betweenness centrality. However, due to the imbalance of
load distribution and the high density of optimal paths, the final
state where the system recovers to is highly unpredictable
\cite{WLL:PREPRINT}. Therefore we are interested in whether these
quantities can return to their \textquotedblleft
normal\textquotedblright , or the steady-state values before the
perturbation.

\begin{figure}[tbp]
\begin{center}
\epsfig{figure=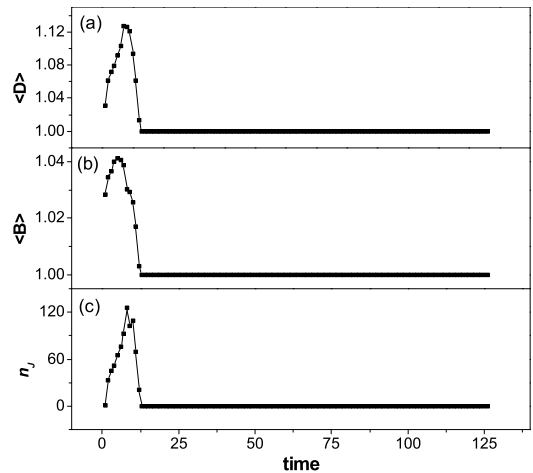,width=0.8\linewidth}
\end{center}
\caption{For a scale-free network of $1000$ nodes and for $\protect\alpha =
0.4 $, time evolutions of three macroscopic quantities: (a) the normalized
network diameter, (b) the normalized betweenness centrality, and (c) the
number of jammed nodes. The quantities return to their respective
steady-state values after a brief transient.}
\label{fig:period_1}
\end{figure}

\begin{figure}[tbp]
\begin{center}
\epsfig{figure=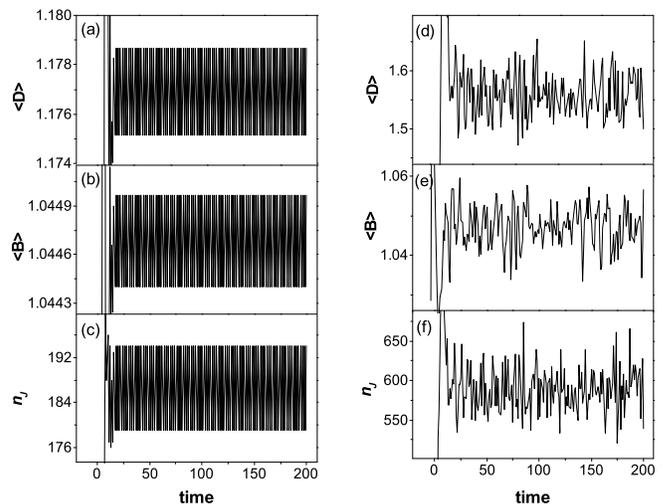,width=\linewidth}
\end{center}
\caption{(a-c) For the same scale-free network but for $\protect\alpha =
0.31 $, periodic oscillations of the normalized diameter (a), the normalized
betweenness centrality (b), and the number of jammed nodes (c). (d-f) random
oscillations of the same set of quantities for $\protect\alpha = 0.2$.}
\label{fig:oscillation}
\end{figure}

For relatively large value of $\alpha $, the ability of the network to
process and transport packets is strong, so we expect the network to be able
to relax to its unperturbed state. This is exemplified in Figs. \ref%
{fig:period_1}(a-c), the time evolutions of three macroscopic quantities,
the normalized diameter $\langle D\rangle $, the normalized betweenness
centrality $\langle B\rangle $, and the number of jammed nodes $n_{J}$,
respectively, of a scale-free network of $1000$ nodes for $\alpha =0.4$.
[For this network the values of the diameter and of the betweenness
centrality in the unperturbed state are $\langle D_{0}\rangle \approx 5.18$
and $\langle B_{0}\rangle \approx 2.35\times 10^{6}$. The plotted quantities
in (a) and (b) are normalized with respect to these \textquotedblleft
static\textquotedblright\ values.] We see that, after about $7$ time steps,
these quantities reach their maximum values and, after another about $5$
steps, these quantities return to their respective unperturbed values. In
this case, the large perturbation causes the network to oscillate but only
for a transient time period. As the capacity parameter $\alpha $ is reduced,
a remarkable phenomenon occurs: after an initial transient the network never
returns to its steady-state but instead, it exhibits persistent
oscillations. Figures \ref{fig:oscillation}(a-c) show periodic oscillations
for $\alpha =0.31$, where the legends are the same as for Figs. \ref%
{fig:period_1}(a-c), respectively. The oscillations are in fact period-2 in
that each macroscopic quantity can assume two distinct values, neither being
the steady-state value, and the quantity alternates between the two values.
For smaller value of $\alpha $, random oscillations \cite%
{Random_oscillations} occur, as shown in Figs. \ref{fig:oscillation}(d-f)
for $\alpha =0.2$.

\begin{figure}[tbp]
\begin{center}
\epsfig{figure=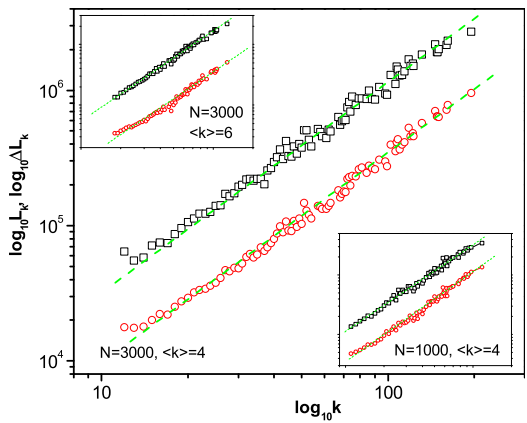,width=0.8\linewidth}
\end{center}
\caption{(Color online) For a scale-free network of $N=3000$ and $\langle
k\rangle =4$, algebraic scaling of the initial load with the degree variable
$k$ (upper data set) and the scaling of the load change caused by unit
weight change (lower data set). The parallelism of the two sets validates
the use of Eq. (\protect\ref{eq:alpha_c}). The insets show similar plots but
for different network parameters. These results suggest that the critical
value $\protect\alpha _{c}$ for network oscillation is independent of the
structural details of the network. The results are averaged over $50$
realizations.}
\label{fig:load}
\end{figure}

The critical value $\alpha _{c}$ of the capacity parameter, below which
persistent network oscillations can occur, can be estimated by noting that,
for a given node $j$, the maximally possible increase in the load before
traffic congestion occurs is $\alpha L_{j}(0)$. The weight-assignment rule
in our traffic protocol, Eq. (\ref{eq:weight_definition}), stipulates that
the most probable weight change be unity. Now regard $\alpha $ as a control
parameter. For a \emph{fixed} amount of change $\Delta L_{j}$ in the load,
free flow of traffic is guaranteed if $\alpha L_{j}(0)>\Delta L_{j}$ but
traffic congestion occurs if $\alpha L_{j}(0)<\Delta L_{j}$. The critical
value $\alpha _{c}$ is then given by
\begin{equation}
\alpha _{c}=\Delta L_{j}/L_{j}(0),  \label{eq:alpha_c}
\end{equation}%
which is independent of the degree variable $k$ \cite{ML:2002}. Since the
load distribution with respect to $k$ is algebraic \cite{GKK:2001}, this
suggests that, in order for Eq. (\ref{eq:alpha_c}) to be meaningful, $\Delta
L_{j}$ must follow an algebraic scaling law with the \emph{same} exponent.
Since the amount of possible weight change is approximately fixed, the
resulting load change is also fixed. To give an example, we consider a
weighted scale-free network of parameters $N=3000$ and $\langle k\rangle =4$%
. Initially all nodes are assigned the same unit weight. The algebraic load
distribution is shown in Fig. \ref{fig:load} (squares, the upper data set)
on a logarithmic scale. The algebraic scaling exponent is about $1.5$. Next
we choose nodes of degree $k$ and give them a sudden, unit increase in the
weight. A recent work shows that for weighted scale-free networks, a weight
increase of a node typically causes its load to decrease \cite{PLY:2004}.
The load change $\Delta L$ as a function of $k$ is shown in Fig. \ref%
{fig:load} (circles, the lower data set). We see that on the logarithmic
scale, $\Delta L$ versus $k$ is parallel to the initial load-degree
distribution curve, justifying the use of Eq. (\ref{eq:alpha_c}).
Numerically we obtain $\alpha _{c}^{\ast }\approx 0.37$. Since in a
realistic situation there are more nodes with weights above the uniform
background value of unity and since the amount of weight change can be more
than unity, this value of $\alpha _{c}$ is only approximate. Indeed, direct
numerical computations give $\alpha _{c}\approx 0.32$. The two estimates are
nonetheless consistent. An interesting observation is that the value of $%
\alpha _{c}$ is insensitive to the network parameters like the network size
and the degree distribution, as shown in the two insets in Fig. \ref%
{fig:load}. In particular, for $N=1000$ and $\langle k\rangle =4$ (inset in
the lower-right corner), we have $\alpha _{c}\approx 0.39$, while for $%
N=3000 $ and $\langle k\rangle =6$ (upper-left corner), we obtain $\alpha
_{c}\approx 0.40$. This phenomenon of network-parameter independency can be
understood by noting that the load variation at a node caused by its weight
change is mainly determined by the probability that optimal paths through
this node appear or disappear, as a result of the weight change. This
probability is independent of the network size and the degree of the node
\cite{PLY:2004}. The value of $\alpha _{c}$, of course, depends on the
weight-assignment rule and thus on the traffic protocol.

\begin{figure}[tbp]
\begin{center}
\epsfig{figure=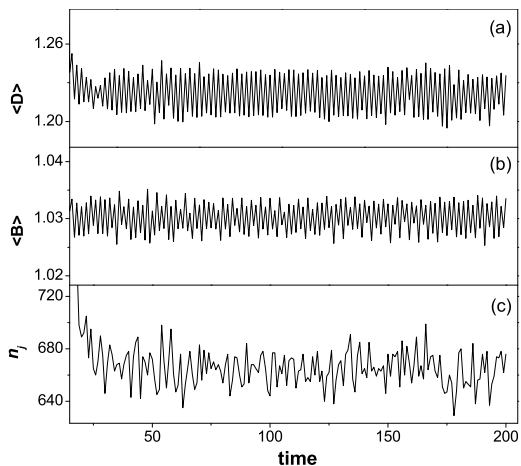,width=0.8\linewidth}
\end{center}
\caption{For the Internet at the autonomous system level with capacity
parameter $\protect\alpha=0.2$, evolutions of the normalized diameter (a),
of the normalized betweenness centrality (b), and of the number of jammed
nodes. Persistent oscillations of the Internet are observed.}
\label{fig:internet}
\end{figure}

Can oscillations be expected in realistic networks? To address this
question, we test the stability of the Internet at the autonomous system
level \cite{MOAT:Web}. The network comprises $6474$ nodes and $13895$ links,
the average diameter is $\langle D(0)\rangle \approx 4.71$, the largest
value of the degree is $1460$ and the load of this node is $L_{j}\approx
1.97\times 10^{8}$. By setting $\alpha =0.2$, we apply perturbation of
strength $P=10\times L_{j}$ at the largest-degree node (to mimic an attack)
and let the Internet evolve according to our traffic protocol. The time
evolutions of the normalized diameter $\left\langle D\right\rangle $, of the
normalized betweenness $\left\langle B\right\rangle $, and of the number of
congested nodes $n_{J}$ are shown in Fig. \ref{fig:internet}(a-c). Again,
persistent oscillations are observed.

Since network oscillation is a result of the dynamical interplay between the
complex topology and the traffic protocol, this phenomenon is expected to be
common for complex systems. For simplicity in this paper we adopt the
standard scale-free random network as the typical model to illustrate the
oscillation, but extensive numerical evidences have shown that this
phenomenon is firmly established for complex networks independent of the
network parameters like the network type, the degree distribution, the
average degree, the network size, the perturbation size and position, and
the specific queuing scheme of the local node dynamics. We note that
although the oscillation phenomenon is robust, its time sequence is very
sensitive to the network details. As a result, the transition from periodic
oscillations to chaotic oscillations as the capacity decreases is
non-smooth. Nonetheless, the trend from regular to complex oscillations is
still clear. To save space, the details will be reported elsewhere \cite%
{WLL:PREPRINT}.

To emphasize the nature of the oscillations, we had avoided the
fluctuations of capacity parameters and traffic load. While these
fluctuations will enhance the oscillations, reflected as larger
oscillation thresholds and larger oscillation amplitudes
\cite{WLL:PREPRINT}, the deterministic feature of the evolutionary
network do exemplify the fact that oscillation is an intrinsic
property of complex systems. This feature makes our model
fundamentally different to the other traffic models in both the
temporal- and spatial-domain \cite{COMPARISON}. Meanwhile, our
model is also different to the congestion control models where
chaotic flux oscillations are observed \cite{VB:2000,GRHA:2005}.
In specific, we consider the competitions of simple node dynamics
on complex topologies while the congestion models investigate the
competitions of complicated node dynamics on simple topologies.

In summary, we have discovered that complex network of finite
capacity can oscillate in the sense that its macroscopic
quantities exhibit persistent periodic or random oscillations in
response to external perturbations. While the study is arisen as a
problem of network security, our findings may have broad
implications to general traffic networks which, when applying the
evolutionary weighted model, need further specifications. Whereas
there can be all sorts of dynamical processes on a complex
network, our findings suggest that there can be physically
meaningful situations where the network itself is never static but
highly dynamic. As a primary model for evolutionary weighted
network, the oscillation of macroscopic quantities is only one
aspect in describing the interplay between the topology and the
local dynamics, further works in other aspects, such as adaptive
networks where the network topology updates according to the local
dynamics, will be very interesting.

YCL thanks the great hospitality of National University of Singapore, where
part of the work was done during a visit. He was also supported by NSF under
Grant No. ITR-0312131 and by AFOSR under Grant No. FA9550-06-1-0024.


\begin{thebibliography}{99}
\bibitem{Optimal_Path} For a non-weighted network, the load on a node is the
total number of shortest paths passing through this node. For a weighted
network, the load depends on the weight distribution and it is more
meaningful to use the optimal path defined with respect to the distribution.
\end{thebibliography}
\end{document}